# Study of the material photon and electron background and the liquid argon detector veto efficiency of the CDEX-10 experiment[*]

SU Jian(苏健)[1]　ZENG Zhi(曾志)[1]　MA Hao(马豪)[1]　YUE Qian(岳骞)[1;1)]　CHENG Jian-Ping(程建平)[1]
CHANG Jian-Ping(常建平)[4]　CHEN Nan(陈楠)[1]　CHEN Ning(陈宁)[1]　CHEN Qing-Hao(陈庆豪)[1]
CHEN Yun-Hua(陈云华)[6]　CHUANG Yo-Chun(庄又澄)[7;2)]　DENG Zhi(邓智)[1]　DU Qiang(杜强)[5]　GONG Hui(宫辉)[1]
HAO Xi-Qing(郝喜庆)[1]　HE Qing-Ju(何庆驹)[1]　HUANG Han-Xiong(黄瀚雄)[2]　HUANG Teng-Rui(黄腾锐)[7;2)]
JIANG Hao(江灏)[1]　KANG Ke-Jun(康克军)[1]　LI Hau-Bin(李浩斌)[7;2)]　LI Jian-Min(李荐民)[1]　LI Jin(李金)[1]
LI Jun(李军)[4]　LI Xia(李霞)[2]　LI Xin-Ying(李新颖)[3]　LI Xue-Qian(李学潜)[3]　LI Yu-Lan(李玉兰)[1]
LI Yuan-Jing(李元景)[1]　LIAO Heng-Yi(廖恒毅)[7;2)]　LIN Fong-Kay(林枫凯)[7;2)]　LIN Shin-Ted(林欣德)[5,7;2)]
LIU Shu-Kui(刘书魁)[5]　LÜ Lan-Chun(吕岚春)[1]　MAO Shao-Ji(毛绍基)[4]　QIN Jian-Qiang(覃建强)[1]　REN Jie(任杰)[2]
REN Jing(任婧)[1]　RUAN Xi-Chao(阮锡超)[2]　SHEN Man-Bin(申满斌)[6]　SINGH Lakhwinder[7,8;2)]
SINGH Manoj Kumar[7,8;2)]　SOMA Arun Kumar[7,8;2)]　TANG Chang-Jian(唐昌建)[5]　TSENG Chao-Hsiung(曾昭雄)[7;2)]
WANG Ji-Min(王继敏)[6]　WANG Li(王力)[5]　WANG Qing(王青)[1]　WONG Tsz-King Henry(王子敬)[7;2)]
WU Shi-Yong(吴世勇)[6]　WU Yu-Cheng(吴玉成)[4]　XING Hao-Yang(幸浩洋)[5]　XU Yin(徐音)[3]　XUE Tao(薛涛)[1]
YANG Li-Tao(杨丽桃)[1]　YANG Song-Wei(杨松纬)[7;2)]　YI Nan(易难)[1]　YU Chun-Xu(喻纯旭)[3]　YU Hao(于昊)[1]
YU Xun-Zhen(余训臻)[5]　ZENG Xiong-Hui(曾雄辉)[6]　ZHANG Lan(张岚)[4]　ZHANG Yun-Hua(张蕴华)[6]
ZHAO Ming-Gang(赵明刚)[3]　ZHAO Wei(赵伟)[1]　ZHOU Zu-Ying(周祖英)[2]　ZHU Jing-Jun(朱敬军)[5]
ZHU Wei-Bin(朱维彬)[4]　ZHU Xue-Zhou(朱雪洲)[1]　ZHU Zhong-Hua(朱忠华)[6]

(CDEX Collaboration)

[1] Key Laboratory of Particle and Radiation Imaging (Ministry of Education) and Department of Engineering Physics, Tsinghua University, Beijing, 100084
[2] Institute of Atomic Energy, Beijing, 102413
[3] Nankai University, Tianjin, 300071
[4] NUCTECH Company, Beijing, 100084
[5] Sichuan University, Chengdu, 610065
[6] Yalong River Hydropower Development Company, Chengdu, 610051
[7] Institute of Physics, Academia Sinica, Taipei, 11529
[8] Department of Physics, Banaras Hindu University, Varanasi, 221005

**Abstract:** The China Dark Matter Experiment (CDEX) is located at the China Jinping underground laboratory (CJPL) and aims to directly detect the WIMP flux with high sensitivity in the low mass region. Here we present a study of the predicted photon and electron backgrounds including the background contribution of the structure materials of the germanium detector, the passive shielding materials, and the intrinsic radioactivity of the liquid argon that serves as an anti-Compton active shielding detector. A detailed geometry is modeled and the background contribution has been simulated based on the measured radioactivities of all possible components within the GEANT4 program. Then the photon and electron background level in the energy region of interest ($<10^{-2}$ events·kg$^{-1}$·day$^{-1}$·keV$^{-1}$ (cpkkd)) is predicted based on Monte Carlo simulations. The simulated result is consistent with the design goal of CDEX-10 experiment, 0.1 cpkkd, which shows that the active and passive shield design of CDEX-10 is effective and feasible.

[*] Supported by the National Natural Science Foundation of China (No.11175099, No.10935005, No.10945002, No.11275107, No.11105076) and National Basic Research Program of China (973 Program) (2010CB833006).
1) Corresponding author. E-mail: yueq@mail.tsinghua.edu.cn
2) Participating as a member of the TEXONO Collaboration.

# 1 Introduction

The CDEX-10 detector is a second generation detector within the China Dark Matter Experiment (CDEX) [1] program, which has 9 p-type point-contact germanium (PCGe) detectors and an active liquid argon (LAr) veto detector. It is dedicated to the direct detection of particle dark matter in the form of weakly interacting massive particles (WIMPs). It is the successor of CDEX-1 [2], which has set some of the best limits on WIMP-nucleon scattering cross sections [3]. The goal for the environmental background counting rate for the CDEX-10 is <0.1 counts per kilogram per keV per day (cpkkd) [4]. To achieve such a low background level, it will be installed in the China Jinping Underground Laboratory (CJPL), which is composed of low radioactivity marble with 2400 m of rock overburden offering an ultra-low background environment with a residual cosmic-ray event rate of $(2.0\pm0.4)\times10^{-10}$ cm$^{-2}\cdot$s$^{-1}$ [5].

The cosmic-ray muons and their progenies that may traverse the shield and contribute to the background are significantly reduced because of the CJPL rock overburden. Thus, the material radioactivity of the shield and the detectors becomes a major factor. The electronic recoil background in the CDEX-10 is from several sources including radioactive contamination in the materials of the shield and the detector and intrinsic radioactivity in the LAr. Even if the veto efficiency is high, dark matter signals may be imitated by the residual electronic recoil events in the nuclear recoil region. To reduce this kind of background, we avoided mounting the radioactive materials close to the germanium crystals.

To achieve a low background level, all materials for the detectors and sub-components have been carefully selected by measurements of radioactivity during the design phase of the CDEX-10. Besides a well-planned design of the passive shield, an active LAr veto detector surrounding the PCGe units further suppresses the background.

In this paper, we summarize the effort to predict the photon and electron background of CDEX-10 from natural radioactivity in shield components and detectors. Background reduction by active veto coincidence cut is studied with Monte Carlo simulations. Section 2 describes the detector model that has been used in the simulations. The transportation of liquid argon scintillation light and the wavelength conversion of optical photons is discussed in Section 3, and the predicted photon and electron background from the detector and shield materials is discussed in Section 4. We draw our conclusions in Section 5.

# 2 CDEX-10 detector design and model simulated with the GEANT4 toolkit

The CDEX-10 detector is a germanium array with a liquid argon veto detector. The total amount of 900 kg of LAr is enclosed in the vacuum insulated cryostat, which is made from the low activity 304SS (stainless steel). The germanium array consists of three PCGe units, each with three 1 kg germanium crystals mounted in the center of the LAr. An array of eleven photomultiplier tubes (PMT) is hung on the top of the LAr to collect scintillation light. To simulate the background of the CDEX-10 detector by photons and electrons and to calculate the LAr veto efficiency, a detailed model (Fig. 1 and Fig. 2) has been created with the GEANT4 toolkit [6].

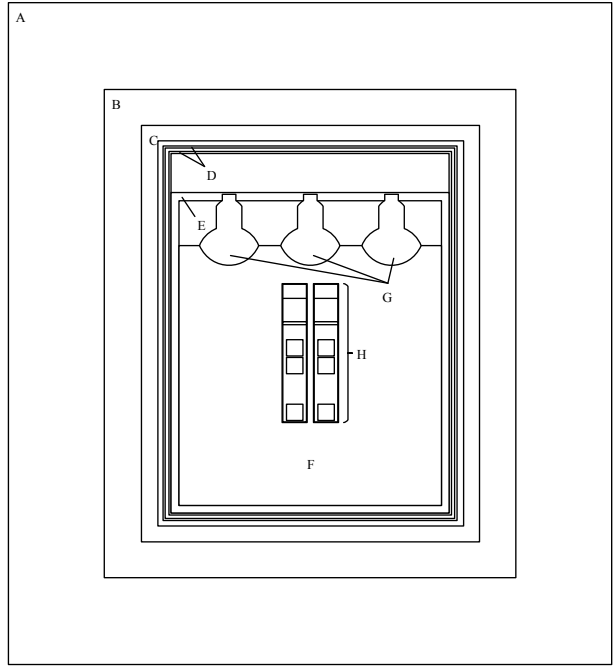

Fig. 1. The GEANT4 model of the CDEX-10 detector and its shield: A—outer polyethylene shield, B—outer lead layer, C—outer copper shield; D—stainless steel cryostat, E—inner copper shield, F—LAr; G—PMT, and H—PCGe unit.

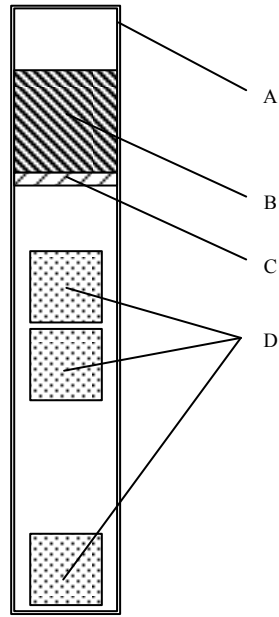

Fig. 2. The GEANT4 model of the PCGe units: A—copper shell, B—lead shield block, C—copper shield block, and D—PCGe. The electronic wires and components in the PCGe units are ignored in the simulation.

Table 1 shows the radioactivity of materials used for the CDEX-10 detector construction, which were computed from the GEANT4 model and are in agreement with the actual detector. The passive lead and copper shield has $4\pi$ detector coverage and is installed in a polyethylene room (thickness 1 m, not shown). From outside to inside, there is a box of 20 cm lead. Inside the lead box, there are 10 cm of copper against the radioactive isotope $^{210}$Pb in a lead box with the other gamma background from the outer shield layers. Highly radioactive components, such as mechanical devices, are located outside the shield.

Table 1. Materials used to construct the CDEX-10 detector, cryostat, and shield as well as their radioactive contamination that were used for Monte Carlo simulations.

| Components | Amounts | Total radioactive contaminations in materials [Bq/amount] | | | |
|---|---|---|---|---|---|
| | | $^{238}$U / $^{226}$Ra | $^{232}$Th | $^{40}$K | other nuclides |
| Polyethylene | 55.0×10$^3$ kg | <412.2×10$^3$ | <483.6×10$^3$ | <2.912×10$^3$ | - |
| Lead shield box | 36.0×10$^3$ kg | 133.6 | 58.5 | 0.984 | $^{210}$Pb: 2.2×10$^6$ |
| Copper shield box | 9.9×10$^3$ kg | 12.2 | 4.01 | 0.374 | - |
| Cryostat (304 SS) | 2.4×10$^3$ kg | 26.4 | 2.89 | 1.37 | $^{60}$Co: 227.7 |
| Copper shield layer | 2.3×10$^3$ kg | 2.89 | 0.948 | 0.0885 | - |
| LAr | 902 kg | - | - | - | $^{39}$Ar: 902.2 |
| Detector copper shell | 3 pieces | 1.21×10$^{-3}$ | 0.365×10$^{-3}$ | 1.39×10$^{-3}$ | - |
| Detector lead shield blocks | 9 pieces | 6.08×10$^{-3}$ | 2.69×10$^{-3}$ | 33.4×10$^{-3}$ | $^{210}$Pb: 6.79 |
| Detector copper shield blocks | 9 pieces | 0.220×10$^{-3}$ | 0.066×10$^{-3}$ | 0.252×10$^{-3}$ | - |
| PMTs | 11 pieces | 4.94 | 1.62 | 24.9 | - |

In addition, the LAr volume around the PCGe units is equipped with PMTs and is an active veto to reduce background by rejecting events in which a particle deposits part of its energy in it.

The cryostat is supported inside the shield. The thickness of the outer cryostat walls is 1.5 mm and the inner is 1.2 mm. The LAr vessel contains a 5 cm thick copper layer to reduce the background from cryostat walls.

The scintillation light is detected by 200 mm (8") diameter 9357UKB PMTs. These are among the lowest radioactivity PMTs and are optimized to operate in the LAr. The top PMT array consists of 11 PMTs mounted in a concentric pattern in a copper support. In the GEANT4 model, each PMT is modeled with two spherical caps and two silica cylinders. The key geometric parameters including the spherical radius, maximum diameter, and waist diameter are the same as the data sheet [7].

## 3 Transportation of liquid argon scintillation light and the wavelength shifting of optical photons

When a particle interacts with the LAr and deposits sufficient energy, 128 nm scintillation light will be emitted isotropically from the interaction point. The scintillation yield is ~40000 photons/MeV. These photons will be converted into the wavelength range (~420 nm) of maximum PMT quantum efficiency by tetraphenyl butadiene (TPB) wavelength shifters (WLS). The TPB is evaporated onto the surface of the copper shell, inner copper shield wall, and PMTs of the PCGe units to wavelength shift the scintillation lights. The 420 nm photons have much longer attenuation length than the 128 nm photons, so they won't be easily absorbed by the LAr and may eventually arrive at the photocathode after several times of reflections.

### 3.1 Simulation of optical photons with wavelength conversion

A LAr scintillation photon is considered optical in GEANT4. In contrast to high energy gamma rays, optical photons are a special class of particles. The processes of optical photons in this simulation include emission of the primary scintillation lights, bulk absorption and Rayleigh scattering, wavelength shifting, reflections, refractions, and PMT quantum efficiencies. The main simulation settings are as follows:
- Scintillation photons

Primary scintillation photons are isotropically emitted in the LAr, with the wavelength distribution ranging from 119 nm to 136 nm [8].

- Bulk absorption

The attenuation length of scintillation photons in LAr is 66 cm [9], which is set as absorption length in the simulation. For the wavelength-shifted optical photons, bulk absorption in LAr is neglected [10], but attenuation length of 20 mm is used in TPB [11].

- Rayleigh scattering

Rayleigh scattering length of 128 nm photons in LAr is set to 90 cm, and for 420 nm photons the length is 400 m [12, 13].

- TPB emission spectrum

Fluorescence photons are emitted from TPB with a wavelength distribution[14]. Even if the TPB thickness or the wavelength of the incident light changes, the emission spectrum of the TPB will not change [11].

- Reflections

The light reflection on the TPB coating is almost completely (>97%) diffuse, irrespective of the TPB coating thickness [15]. The reflection coefficient in the TPB emission spectral region was measured to be is about 95% [10, 16].

- Refractions

The refraction index of LAr at 128 nm is 1.38 [17], and the refraction index of TPB is set to 1.60 for 128 nm [18] and 1.27 for 420 nm [19].

- PMT collection

The quantum efficiency curve [7] of the 9357KB PMT is used to convert the number of photons collected by PMTs to the number of photoelectrons.

## 3.2 Calculation of the LAr veto detector response threshold of energy deposition

According to the simulation process, the calculated value of the total optical photon collection efficiency is 0.05355, which means that each scintillation photon in the LAr will eventually result in 0.05355 photoelectrons collected by the PMTs on average.

Both optical photons and electronic noise may trigger a PMT. If an event in HPGe is vetoed as long as any PMT collects photoelectrons, too many signals may be lost. To reduce the impact of electronic noise on the PMT, we initially set a veto criteria as follows: If two or more PMTs collected ≥ 2 photoelectrons at the same time, the veto was executed. We calculated the probability distribution of the photoelectron collection by the PMTs to determine the energy threshold of the veto criteria.

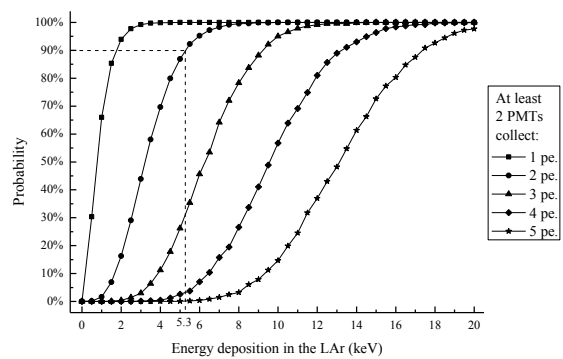

Fig. 3. Calculated probability distribution of the photoelectrons (pe.) collected by the PMTs. The dashed line shows the determination of the energy threshold.

As more energy is deposited in the LAr, the more the PMTs collect photoelectrons (Fig. 3). We adopted the situation of 90% probability and obtained an energy threshold of 5.3 keV from Fig. 3. That means when 5.3 keV energy deposits in the LAr, two or more PMTs will respectively collect 2 photoelectrons with 90% probability. This energy threshold is used as the response threshold of the LAr veto detector in the following simulations.

# 4 Background due to radioactive contamination in shield materials and the detector

Special care has been taken to select shield and detector materials according to their radioactive contaminations. The majority of the materials used in the shield and the detectors were measured with low background HPGe detectors to determine their radioactivities, mainly due to the $^{238}$U series, $^{232}$Th series, $^{40}$K, and $^{60}$Co.

The radioactive contaminations of the materials are shown in Table 1. The radioactive equilibrium of the $^{238}$U chain and the $^{232}$Th chain is assumed. The measured activities of the materials have been used in the Monte Carlo simulations to predict the background.

## 4.1 Veto judgment

Due to the very low scattering cross section of the WIMPs, the predicted WIMP event should be a single scatter event. There are ten detectors in the CDEX-10—nine PCGe detectors and one LAr veto detector. If a particle has deposited enough energy at multiple detectors (0.5 keV for germanium detectors and 5.3 keV for the LAr veto detector), such an event is a multiple scatter event and will be rejected in the following data analysis.

For each event recorded by PCGe in CDEX-10, the number of the triggered detectors is checked. Table 2 shows the judgment basis.

Table 2. Active veto judgment basis

| The number of the triggered PCGe detectors | Whether the LAr is triggered? | Judgments | Accept or reject |
|---|---|---|---|
| = 0 | Yes or No | Not recorded | / |
| ≥ 1 | Yes | Vetoed by the LAr | Reject |
| ≥ 2 | No | Vetoed by PCGe | Reject |
| = 1 | No | Residual events | Accept |

## 4.2 Simulated background spectrum

Fig. 4 shows the predicted spectrum in the entire energy range, and Fig. 5 is in the region of interest. The energy range for the background rate calculation is up to 100 keV to include the signal region for inelastic dark matter from standard elastic WIMP scattering.

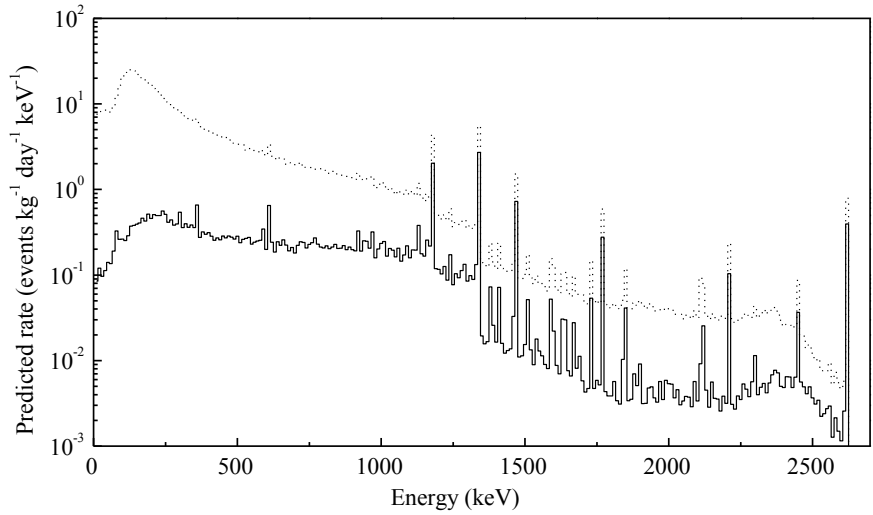

Fig. 4. Predicted background from the detector and shield materials: energy spectra of all events (dot line) and residual events (solid line) in the entire energy range.

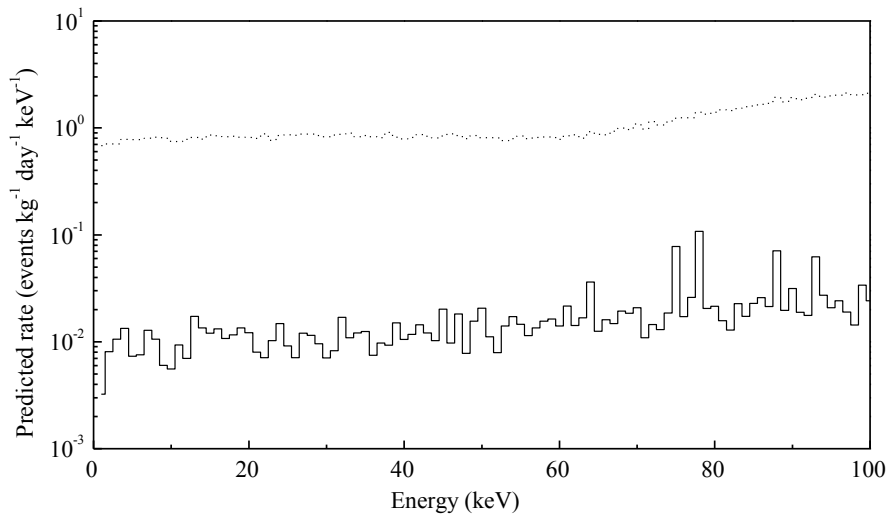

Fig. 5. Zoom into the low energy region of the Monte Carlo spectra shown in Fig. 2. The spectra of all events (dot line) and residual events (solid line) in the energy range up to 100 keV.

### 4.3 Background count rates and active veto efficiency

By considering all possible effects in the design, the target physical event rate of CDEX-10 is 0.1 cpkkd [1]. From the simulated results in Table 3 and Table 4, the total event rates with an active veto of 0.5~2700 keV and 0.5~10 keV are 0.0154 cpkkd and 0.0085 cpkkd, respectively.

Table 3. Predicted rates in the energy range of 0.5~2700 keV from the components of CDEX-10. Data are presented descending order according to the rates of active veto.

| Components | Predicted rates (cpkkd) | | Veto efficiencies | Component location types |
|---|---|---|---|---|
| | active veto | none veto | | |
| SS Cryostat (inner wall) | $5.47\times10^{-3}$ | $8.18\times10^{-2}$ | 93.32% | outside the LAr |
| SS Cryostat (outer wall) | $5.44\times10^{-3}$ | $7.98\times10^{-2}$ | 93.17% | outside the LAr |
| PMTs | $1.80\times10^{-3}$ | $4.30\times10^{-2}$ | 95.81% | outside the LAr |
| Detector copper shell | $1.75\times10^{-3}$ | $3.39\times10^{-3}$ | 48.45% | inside the LAr |
| Copper shield layer | $1.09\times10^{-3}$ | $1.71\times10^{-2}$ | 93.63% | outside the LAr |
| Detector lead shield blocks | $7.00\times10^{-4}$ | $1.76\times10^{-3}$ | 60.31% | inside the LAr |
| Detector copper shield blocks | $1.23\times10^{-4}$ | $2.63\times10^{-4}$ | 53.19% | inside the LAr |
| Copper shield box | $5.35\times10^{-5}$ | $8.16\times10^{-4}$ | 93.44% | outside the LAr |
| LAr | $3.53\times10^{-5}$ | $2.39\times10^{-2}$ | 99.85% | LAr itself |
| Lead shield box | $2.41\times10^{-7}$ | $8.19\times10^{-6}$ | 97.06% | outside the LAr |
| Total | 0.0165 | 0.2518 | 93.46% | |

Table 4. Predicted rates in the energy range of 0.5~10 keV from the components of CDEX-10. Presented in descending order according to the rates of active veto.

| Components | Predicted rates (cpkkd) | | Veto efficiencies | Component location types |
|---|---|---|---|---|
| | active veto | none veto | | |
| Detector copper shell | $2.99\times10^{-3}$ | $1.14\times10^{-2}$ | 73.87% | inside the LAr |
| SS Cryostat (inner wall) | $2.76\times10^{-3}$ | $2.67\times10^{-1}$ | 98.97% | outside the LAr |
| SS Cryostat (outer wall) | $1.46\times10^{-3}$ | $2.59\times10^{-1}$ | 99.44% | outside the LAr |
| PMTs | $6.17\times10^{-4}$ | $1.35\times10^{-1}$ | 99.54% | outside the LAr |
| Detector lead shield blocks | $5.83\times10^{-4}$ | $6.39\times10^{-3}$ | 90.88% | inside the LAr |
| Copper shield layer | $4.95\times10^{-4}$ | $5.34\times10^{-2}$ | 99.07% | outside the LAr |
| Detector copper shield blocks | $1.60\times10^{-4}$ | $9.07\times10^{-4}$ | 82.38% | inside the LAr |
| Copper shield box | $3.97\times10^{-5}$ | $2.39\times10^{-3}$ | 98.34% | outside the LAr |
| LAr | $<1\times10^{-5}$ | $7.19\times10^{-2}$ | ~100% | LAr itself |
| Lead shield box | $<1\times10^{-9}$ | $1.26\times10^{-7}$ | ~100% | outside the LAr |
| Total | 0.0091 | 0.8081 | 98.87% | |

For the PCGe, detector units are entirely surrounded by the LAr. The component locations can be divided into three types: "LAr itself", "outside the LAr", and "inside the LAr". Table 3 and Table 4 show that the veto efficiency mainly depends on the location of the components, which is "LAr itself" > "outside the LAr" > "inside the LAr". The reasons are:

1. In the LAr itself, $^{39}$Ar emits electrons that certainly deposit energy in the LAr and be mostly vetoed. Only a small fraction of the electrons emitted very close to PCGe shell may not trigger the LAr and emit bremsstrahlung photons into PCGe with low probability. Thus, the veto efficiency is ~100%.

2. From outside the LAr, the photons must pass through 37 cm thick LAr without reaction, and these events will not be vetoed. However, this probability is quite low, and the veto efficiency is >93% (0.5~2700 keV) and >98% (0.5~10 keV).

3. From inside the LAr, the particles can easily hit the PCGe directly without energy deposition in the LAr. Only a part of the particles may scatter into the LAr after hitting the PCGe. Thus, the veto efficiency is 48%~60% (0.5~2700 keV) and 73%~90% (0.5~10 keV).

## 5 Conclusion

A study to predict the photon and electron background of the CDEX-10 experiment has been performed. Monte Carlo simulations are performed with GEANT4 by using the measured radioactivity values of all the relevant components and establishing a detailed geometry model of the shield and detector.

Taking the processes of optical photons into consideration, a value of 0.05355 for total optical photon collection efficiency was calculated in the simulation. These processes include emission of the primary scintillation lights, bulk absorption, Rayleigh scattering, wavelength shifting, reflections, refractions, and PMT quantum efficiencies. The response threshold of the LAr veto detector is assumed to be 5.3 keV.

The predicted rate of the photon and electron background events in the energy region below 10 keV with veto coincidence cut is 0.7530 events·kg$^{-1}$·day$^{-1}$·keV$^{-1}$ for the 9 kg germanium mass. By applying a veto cut with an energy threshold of 5.3 keV for the LAr veto detector, these rates are reduced to 0.0084 events·kg$^{-1}$·day$^{-1}$·keV$^{-1}$. The veto efficiencies mainly depend on the location of the components, which are "LAr itself" (~100%) > "outside the LAr" (98.3~99.5%) > "inside the LAr" (73%~90%) in the energy range of 0.5~10 keV.

According to the simulated results in this paper, the design goal of CDEX-10—0.1 events·kg$^{-1}$·day$^{-1}$·keV$^{-1}$—may indeed be achieved. This is due to the materials selection, an improved passive shield, and the use of an active LAr veto. The results of this work are not only important for understanding the photon and electron background in the CDEX-10 experiment and the validation of the background model, but are also useful for the design of next generation detectors for dark matter searches such as the CDEX-1T [1].